\documentclass[10.5pt, conference, letterpaper]{IEEEtran}
\IEEEoverridecommandlockouts
  \usepackage{caption}
  \usepackage[table,xcdraw]{xcolor}
  \usepackage{comment}
  \usepackage{lipsum}
  
\usepackage{cite}
\usepackage{amsmath,amssymb,amsfonts}
\usepackage{algorithmic}
\usepackage{algorithm}
\usepackage[applemac]{inputenc} 
\newcommand{\RN}[1]{%
  \textup{\uppercase\expandafter{\romannumeral#1}}%
}
\newcommand*\sepline{%
   \begin{center}
     \rule[1ex]{0.9\textwidth}{.4pt}
   \end{center}}
\usepackage{float}
\usepackage{cuted,nccmath}
\usepackage{cite}
\usepackage{amsmath,amssymb,amsfonts}
\usepackage{graphicx}
\usepackage{textcomp}
\usepackage{xcolor}
\def\BibTeX{{\rm B\kern-.05em{\sc i\kern-.025em b}\kern-.08em
    T\kern-.1667em\lower.7ex\hbox{E}\kern-.125emX}}
\begin{document}


\title{Securing V2I Communications in 5G and Beyond Wireless System Using Gradient Ascent Approach}

\author{Neji~Mensi,~\IEEEmembership{Member,~IEEE,} Danda~B.~Rawat,~\IEEEmembership{Senior Member,~IEEE,} and Elyes~Balti,~\IEEEmembership{Member,~IEEE}
\thanks{This work was supported in part by the US NSF under grants CNS 1650831.}
\thanks{
Neji Mensi and Danda B. Rawat are with the Department
of Electrical Engineering and Computer Science, Howard University, Washington, DC, 20059 USA e-mail: neji.mensi@bison.howard.edu, danda.rawat@howard.edu.}
\thanks{Elyes Balti is with the Wireless Networking and Communications Group, Department of Electrical and Computer Engineering, The University
of Texas at Austin, Austin, TX 78712 USA e-mail: ebalti@utexas.edu.}
}

\maketitle

\begin{abstract}
The 5G and beyond wireless systems, instead of being just an extension of 4G, are  regarded as `network of networks' which is expected to integrate heterogeneous wireless networks including wireless vehicular networks (WVN). 
The WVN promises to solve many issues such as reducing road accidents, traffic jams, fuel consumption and commute time, which  is subject to various security issues such as eavesdropping, where attackers attempt to overhear the secret transmitted signal passively (which makes the detection/defense very difficult).  In this work, our objective is to implement a security scheme by improving the secrecy capacity in Sub-6 GHz and millimeter-wave (mmWave) bands.  Vehicle-to-Infrastructure (V2I) for 5G and beyond wireless network in . Unlike previous works where the secrecy capacity is increased by randomly augmenting the source power or setting the system combiners/precoders, in this work we employ the  Gradient Ascent algorithm to search for the best combiners/precoders that maximizes the secrecy rate performance. We further present two different optimization scenarios: fixed and variable transmission power.
\end{abstract}


\section{Introduction}
\subsection{Background and Literature Review}
Vehicular network promises to solve many issues such as reducing road accidents, traffic jams, fuel consumption and commute time \cite{tayyaba20205g} that make it grow  exponentially. 
The microwave band is very limited and incapable to support such advanced network technology. Fortunately, Sub-6 GHz and millimeter-wave (mmWave) technologies have been considered   promising solutions to address the aforementioned constraints
\cite{mmw}. 
Though, despite the high data rate and its highly advanced services,  wireless vehicular network (WVN) still vulnerable to various security menaces, which may seriously affect its spread among regular users \cite{arif2019survey}. The eavesdropping attack is one of the most severe cyber threats. In such attack, the eavesdropper attempts to listen to the signal exchanged between authorized nodes. It is a critical attack, especially if the intercepted signal is classified or if the disclosing of the secret message may influence the communicating entities \cite{generalSec1}. 
To confront this challenge, physical layer security (PLS) has been discussed and introduced as an emerging security field that mitigates such passive attack. The main idea of PLS is to take advantage of the physical layer's properties such as the channels' characteristics of wireless communication\cite{PLS1}.
One of the PLS techniques is employing a jammer, where its main function is to jam and perturb the eavesdropper link \cite{myJam}. Thereby, the signal-to-interference-plus-noise ratio (SINR) at the eavesdropper end will be decreased. This technique improves the average secrecy capacity and protects the communications between the legitimate nodes significantly. 

\subsection{Our Contribution}
The setting of the optimum transmission power that will increase the secrecy capacity may be challenging. Therefore, in this paper, we suggest the use of machine learning to enhance security while considering a fully analog beamforming design. We are focusing on adapting a friendly jammer to protect Vehicle-to-Infrastructure (V2I) communications while manipulating the precoders at the sources and the combiners at the receivers to increase the secrecy capacity.  To the best of the authors' knowledge, unlike the previous work in the literature,  this is the first attempt to maximize the secrecy capacity using the Gradient Ascent approach for the V2I communications in  5G network while considering a friendly jammer and the scenarios of fixed/variable transmission power. 
We highlight our contributions as follows:
\begin{enumerate}
\item[$\bullet$] Studying the PLS for two different scenarios: Sub-6 GHz and mmWave systems.
  \item[$\bullet$] Develop a Gradient Ascent algorithm to maximize the secrecy capacity for fixed source power.
   \item[$\bullet$] Develop a second Gradient Ascent algorithm to reach a desirable secrecy capacity while adapting the source power. 
\end{enumerate}
\subsection{Paper Structure}
The paper is organized as follows:  Section \RN{2} presents the system and models. Section \RN{3} studies adaptive Gradient ascent constraints and its formulas derivations.  Then, Section \RN{4} assesses the performance of the security strategy and the algorithm convergences, based on numerical results, for Sub-6 GHz and mmWave systems. Finally, we outline our conclusion in Section \RN{5}.
\subsection{Notation}
For the purpose of simplicity and organization, we utilize the following notations: $||\cdot||_{2}$ is the 2-norm, $|\cdot|$ is the absolute value, and  $(\cdot)^{*}$ is the Hermitian operator,

\section{System Model}
\begin{figure*}[ht]
\includegraphics[height=8cm, width=\linewidth ]{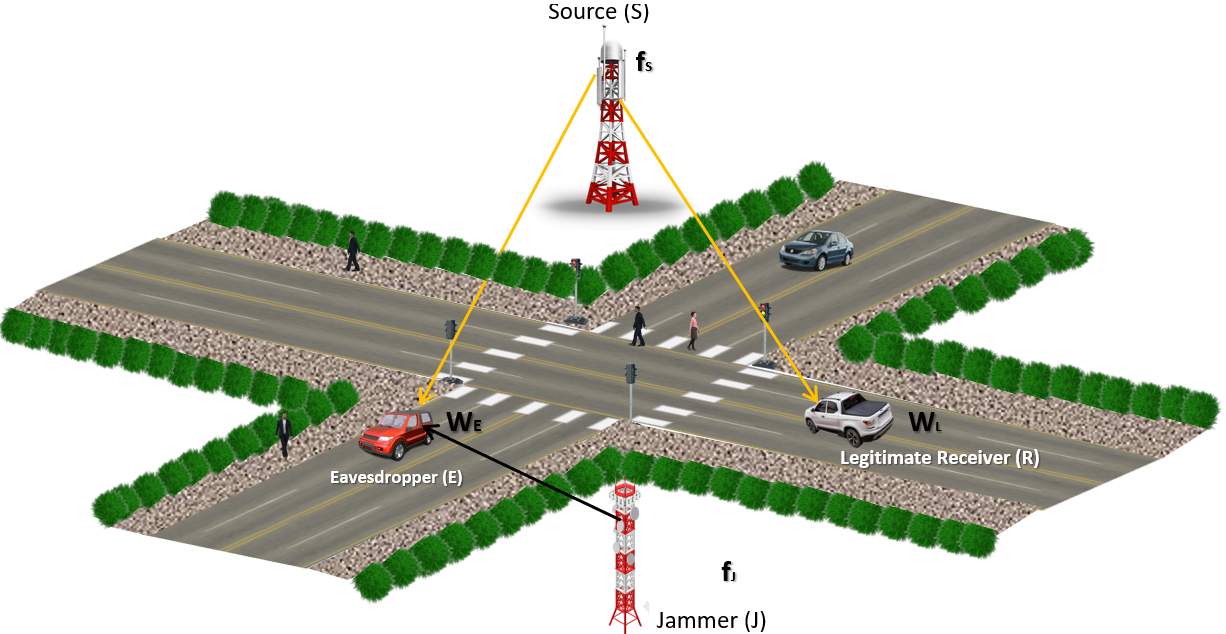}
\caption{V2I communications in the presence of an eavesdropper.  }
\end{figure*}
\subsection{Vehicular Communications and Attack Model}
The vehicular network needs to satisfy a high quality of service (QoS), especially security.  Therefore, the signals exchanged between BS and vehicles in V2I communication should be protected and kept confidential.
Eavesdropping is a crucial threat in such communications scenarios,
where the attacker intercepts the confidential signal.
The communication model in Fig. 1 presents a typical model where the attacker \textit{E}  is overhearing the communication between the base station (BS) noted by $S$ and the legitimate receiver $L$. To jam the signal received by the eavesdropper, we intend to employ a jammer $J$, where its main goal is perturbing and minimizing the SINR at the attacker's end. \\
We note that $S$ and $J$ are equipped with $N_T$ antennas, and both the legitimate receiver and the attacker are equipped with $N_R$ antennas.

\subsection{Channel Model}
In this paper, we assume a Sub-6 GHz/mmWave system. Therefore, the channels follow a geometric model based on clusters and rays:
\begin{equation}
H= \sqrt{\frac{N_R N_T}{N_{cl}N_{ray}}}\sum_{i=1}^{N_{cl}}\sum_{j=1}^{N_{ray}}\beta_{i,j}a_{R}(\phi_{i,j}^{a},\theta_{i,j}^{a})a_{T}^{*}(\phi_{i,j}^{d},\theta_{i,j}^{d})
\end{equation}
where $N_{cl},N_{ray}, N_R,$ and $N_T$ are, respectively, the number of clusters, the number of rays, the number of antennas at the receiver, and antennas at the transmitter. $\beta_{i,j}$ is the complex gain of the $i^{th}$ cluster and the $j^{th}$.
ray. The array steering and response vectors at the receiver (R) and the transmitter (T) are $a_{R}(\phi_{i,j},\theta_{i,j})$ and $a_{T}^{*}(\phi_{i,j},\theta_{i,j})$, respectively. The angles of departure (DoD) and the angles of arrival (AoA) are denoted by 
$\phi_{i,j}^{d},\theta_{i,j}^{d}$ and $\phi_{i,j}^{a},\theta_{i,j}^{a}$, respectively

The received signal at \textit{L} and \textit{E}  are respectively:
\begin{equation}
\begin{aligned}
{ y_{_L}}& = \textbf{w}^{*}_{_L}\textbf{H}_{sl}\textbf{f}_{s}s_{s} + \textbf{w}^{*}_{_L}\textbf{H}_{jl}\textbf{f}_{j}s_{j} + \textbf{n}_{_L}
\end{aligned}
\end{equation}
\begin{equation}
\begin{aligned}
{ y_{_E}}& = \textbf{w}^{*}_{_E}\textbf{H}_{se}\textbf{f}_{s}s_{j} + \textbf{w}^{*}_{_E}\textbf{H}_{je}\textbf{f}_{j}s  + \textbf{n}_{_E}
\end{aligned}
\end{equation}
where $\textbf{w}_{a}$ is  the combiner vector at the node $a$ with dimension $N_{R}\times 1$ ($a \in\{L,E\}$). $\textbf{H}_{sl}, \textbf{H}_{se}$, $\textbf{H}_{jl}$ and $\textbf{H}_{je}$ are respectively, the channel matrix from $S$ to $L$, $S$ to $E$,  $J$ to $L$, and $J$ to $E$ with dimensions $N_{R} \times N_{T}$. The precoders applied at $J$ and $S$ are defined by $f_s$ and $f_j$ with dimension $N_{T} \times 1$.  $s_{b}$ is the information symbol sent by transmitter $b$ where $b \in \{S,J\}$. Finally, $n_{a}$ is the zero mean additive white Gaussian noise (AWGN)  with variance $\sigma_{a}$.

The received SNR and SINR at $R$ and $E$, respectively, are
\begin{equation}
\begin{aligned}
{ \gamma_{_L}}& = \frac{P_{s} |\textbf{w}^{*}_{_L}\textbf{H}_{sl}\textbf{f}_{s}|^2}{\textbf{w}_{_L}^{*}\textbf{w}_{_L}\sigma_{_L}^2 + P_{j} |\textbf{w}^{*}_{_L}\textbf{H}_{jl}\textbf{f}_{j}|^2}\\
&=\frac{P_{s} (\textbf{w}^{*}_{_L}\textbf{H}_{sl}\textbf{f}_{s})(\textbf{w}^{*}_{_L}\textbf{H}_{sl}\textbf{f}_{s})^{*}}{\textbf{w}_{_L}^{*}\textbf{w}_{_L}\sigma_{_L}^2 + P_{j} (\textbf{w}^{*}_{_L}\textbf{H}_{jl}\textbf{f}_{j})(\textbf{w}^{*}_{_L}\textbf{H}_{jl}\textbf{f}_{j})^{*} }.
\end{aligned}
\end{equation}
\begin{equation}
\begin{aligned}
{ \gamma_{_E}}& = \frac{P_{s} |\textbf{w}^{*}_{_E}\textbf{H}_{se}\textbf{f}_{s}|^2}{\textbf{w}_{_E}^{*}\textbf{w}_{_E}\sigma_{_E}^{2} + P_{j} |\textbf{w}^{*}_{_E}\textbf{H}_{je}\textbf{f}_{j}|^2 }\\
&=\frac{P_{s} (\textbf{w}^{*}_{_E}\textbf{H}_{se}\textbf{f}_{s})(\textbf{w}^{*}_{_E}\textbf{H}_{se}\textbf{f}_{s})^{*}}{\textbf{w}_{_E}^{*}\textbf{w}_{_E}\sigma_{_E}^{2} + P_{j} (\textbf{w}^{*}_{_E}\textbf{H}_{je}\textbf{f}_{j})(\textbf{w}^{*}_{_E}\textbf{H}_{je}\textbf{f}_{j})^{*} }.
\end{aligned}
\end{equation}

To improve the security performance of our communications model, we need to maximize the average secrecy capacity $C_{_s}$, which is defined as follows
\begin{equation}
    C_{_s}=\max[(C_{_L}-C_{_E}),0], \label{costf}
\end{equation} 
where $C_{_L}$ and $C_{_E}$ are respectively the ergodic capacities at $L$ and $E$ and they are defined by
\begin{equation}
\begin{aligned}
{ C_{_L}}& = \log_2 \left( 1+ \frac{P_{s} (\textbf{w}^{*}_{_L}\textbf{H}_{sl}\textbf{f}_{s})(\textbf{w}^{*}_{_L}\textbf{H}_{sl}\textbf{f}_{s})^{*}}{\textbf{w}_{_L}^{*}\textbf{w}_{_L}\sigma_{_L}^2 + P_{j} (\textbf{w}^{*}_{_L}\textbf{H}_{jl}\textbf{f}_{j})(\textbf{w}^{*}_{_L}\textbf{H}_{jl}\textbf{f}_{j})^{*} }\right).
\end{aligned}
\end{equation}

\begin{equation}
\begin{aligned}
{ C_{_E}}& = \log_2 \left ( 1 + \frac{P_{s} (\textbf{w}^{*}_{_E}\textbf{H}_{se}\textbf{f}_{s})(\textbf{w}^{*}_{_E}\textbf{H}_{se}\textbf{f}_{s})^{*}}{\textbf{w}_{_E}^{*}\textbf{w}_{_E}\sigma_{_E}^{2} + P_{j} (\textbf{w}^{*}_{_E}\textbf{H}_{je}\textbf{f}_{j})(\textbf{w}^{*}_{_E}\textbf{H}_{je}\textbf{f}_{j})^{*} } \right ).
\end{aligned}
\end{equation}
\section{Adaptive Gradient Ascent Algorithm}
 \subsection{Gradient Learning Approach Without Optimizing $W_{_E}$}
Our goal is to maximize the secrecy capacity by searching for the adequate  values of $\textbf{w}_{_L}, \textbf{f}_{s}$, and $\textbf{f}_{j}$. Therefore, we are going to maximize the cost function in Eq. (\ref{costf}) under two constraints: the  unit-norm and the CA constraints.
We can formulate the optimization problem as follows
\begin{equation}\label{maxrate}
\begin{split}
\mathcal{P}: \max\limits_{\bold{w}_{_L},\bold{f}_s, \bold{f}_j}& C_{s}
\end{split}
\end{equation}
\begin{equation}\label{c1}
\text{subject to}~~|\bold{f}_s| = |\bold{f}_j| = |\bold{w}_{_L}| =  1
\end{equation}
\begin{equation}\label{c2}
~~~~~~~~~~~~~~~~~~~~~~~~~~~~\bold{w}_{_L} \in\mathcal{G}^{N_R} ~\text{and}~ \bold{f}_s, \bold{f}_j \in\mathcal{G}^{N_T}
\end{equation}

where Eq. (\ref{c1}) is the unit-norm constraint and Eq. (\ref{c2}) is the CA constraint (\begin{math}\mathcal{G}^{N_T}\end{math} and 
\begin{math}\mathcal{G}^{N_R}\end{math} are the subspace of the CA constraints of dimensions $N_T$ and $N_R$, respectively)\cite{mmwArx}.
We need to compute the gradients with respect to $\textbf{w}_{_L}, \textbf{f}_{s}$, and $\textbf{f}_{j}$, then project the solutions onto the subspace of the CA constraints. Therefore, we will start by differentiating the cost function for $\textbf{w}_{_L}$, which is expressed as follow

\begin{equation}
\begin{aligned}
&\nabla_{\textbf{w}_{_L}^{*}}C_{_s} =\frac{\partial C_{_s}}{\partial\textbf{w}_{_L}^{*}}= \left[ \frac{\partial C_{_L}}{\partial\textbf{w}_{_L}^{*}}-\frac{\partial C_{_E}}{\partial\textbf{w}_{_L}^{*}} \right]=\frac{\partial C_{_L}}{\partial\textbf{w}_{_L}^{*}}\\
&=\frac{\textbf{w}_{_L}\sigma_{_L}^{2} + P_{j}\textbf{H}_{jl}\textbf{f}_{j}\textbf{f}_{j}^{*}\textbf{H}_{jl}^{*}\textbf{w}_{_L} +P_{s}\textbf{H}_{sl}\textbf{f}_{s}\textbf{f}_{s}^{*}\textbf{H}_{sl}^{*}\textbf{w}_{_L}}{\sigma_{_L}^{2} + P_{j}\Psi_{jl}+P_{s}\Psi_{sl} }\\
&-\frac{\textbf{w}_{_L}\sigma^2 + P_{j}\textbf{H}_{jl}\textbf{f}_{j}\textbf{f}_{j}^{*}\textbf{H}_{jl}^{*}\textbf{w}_{_L} }{\sigma_{_L}^{2} + P_{j} \Psi_{jl}}
\end{aligned}
\end{equation}
where $\Psi_{jl}=(\textbf{w}^{*}_{_L}\textbf{H}_{jl}\textbf{f}_{j})(\textbf{w}^{*}_{_L}\textbf{H}_{jl}\textbf{f}_{j})^{*}$ and $\Psi_{sl}=(\textbf{w}^{*}_{_L}\textbf{H}_{sl}\textbf{f}_{s})(\textbf{w}^{*}_{_L}\textbf{H}_{sl}\textbf{f}_{s})^{*}$.\\
As a second step, we compute the gradient of $\textbf{f}_{j}$, which is given by
\begin{equation}
\begin{aligned}
&\nabla_{\textbf{f}_{j}^{*}}C_{_s} =\frac{\partial C_{_s}}{\partial\textbf{f}_{j}^{*}}= \left[ \frac{\partial C_{_L}}{\partial\textbf{f}_{j}^{*}}-\frac{\partial C_{_E}}{\partial\textbf{f}_{j}^{*}} \right]
\end{aligned}
\end{equation}
where
\begin{equation}
\begin{aligned}
\frac{\partial C_{_L}}{\partial\textbf{f}_{j}^{*}} &=
\frac{P_{j}\textbf{H}_{jl}^{*}\textbf{w}_{_L}\textbf{w}_{_L}^{*}\textbf{H}_{jl}\textbf{f}_{j}}{\sigma_{_L}^{2} + P_{j} \Psi_{jl} + P_{s} \Psi_{sl}}
-\frac{P_{j}\textbf{H}_{jl}^{*}\textbf{w}_{_L}\textbf{w}_{_L}^{*}\textbf{H}_{jl}\textbf{f}_{j}}{\sigma_{_L}^{2} + P_{j} \Psi_{jl} }
\end{aligned}
\end{equation}
\begin{equation}
\begin{aligned}
\frac{\partial C_{_E}}{\partial\textbf{f}_{j}^{*}} &=
\frac{P_{j}\textbf{H}_{je}^{*}\textbf{w}_{_E}\textbf{w}_{_E}^{*}\textbf{H}_{je}\textbf{f}_{j}}{\sigma_{_E}^{2} + P_{j} \Psi_{je} + P_{s} \Psi_{se}}
-\frac{P_{j}\textbf{H}_{je}^{*}\textbf{w}_{_E}\textbf{w}_{_E}^{*}\textbf{H}_{je}\textbf{f}_{j}}{\sigma_{_E}^{2} + P_{j} \Psi_{je} }
\end{aligned}
\end{equation}
where $\Psi_{je}=(\textbf{w}^{*}_{_E}\textbf{H}_{je}\textbf{f}_{j})(\textbf{w}^{*}_{_E}\textbf{H}_{je}\textbf{f}_{j})^{*}$ and $\Psi_{se}=(\textbf{w}^{*}_{_E}\textbf{H}_{se}\textbf{f}_{s})(\textbf{w}^{*}_{_E}\textbf{H}_{se}\textbf{f}_{s})^{*}$.\\

\begin{figure*}
\begin{equation}\label{gradientFj}
\begin{aligned}
\nabla_{\textbf{f}_{j}^{*}}C_{_s}=\frac{P_{j}\textbf{H}_{jl}^{*}\textbf{w}_{_L}\textbf{w}_{_L}^{*}\textbf{H}_{jl}\textbf{f}_{j}}{\sigma_{_L}^{2} + P_{j} \Psi_{jl} + P_{s} \Psi_{sl}}
-\frac{P_{j}\textbf{H}_{jl}^{*}\textbf{w}_{_L}\textbf{w}_{_L}^{*}\textbf{H}_{jl}\textbf{f}_{j}}{\sigma_{_L}^{2} + P_{j} \Psi_{jl} }-
\frac{P_{j}\textbf{H}_{je}^{*}\textbf{w}_{_E}\textbf{w}_{_E}^{*}\textbf{H}_{je}\textbf{f}_{j}}{\sigma_{_E}^{2} + P_{j} \Psi_{je} + P_{s} \Psi_{se}}
+\frac{P_{j}\textbf{H}_{je}^{*}\textbf{w}_{_E}\textbf{w}_{_E}^{*}\textbf{H}_{je}\textbf{f}_{j}}{\sigma_{_E}^{2} + P_{j} \Psi_{je} }.
\end{aligned}
\end{equation}
\sepline
\end{figure*}
Therefore, the gradient of $\textbf{f}_{j}$ is expressed by Eq. (\ref{gradientFj}) on the top of the next page.
Finally, we derive the gradient  of $\textbf{f}_{s}$ as follow
\begin{equation}
\begin{aligned}
\nabla_{\textbf{f}_{s}^{*}}C_{_s}& =\frac{\partial C_{_s}}{\partial\textbf{f}_{s}^{*}}= \left[ \frac{\partial C_{_L}}{\partial\textbf{f}_{s}^{*}}-\frac{\partial C_{_E}}{\partial\textbf{f}_{s}^{*}} \right].
\end{aligned}
\end{equation}
Hence, we can first compute the partial derivative of the rate at the legitimate receiver node:
\begin{equation}
\begin{aligned}
\frac{\partial C_{_L}}{\partial\textbf{f}_{s}^{*}} &=
\frac{P_{j}\textbf{H}_{jl}^{*}\textbf{w}_{_L}\textbf{w}_{_L}^{*}\textbf{H}_{jl}\textbf{f}_{j}}{\sigma_{_L}^{2} + P_{j} \Psi_{jl} + P_{s} \Psi_{sl}}.
\end{aligned}
\end{equation}
Then, we  derive the partial derivative of the rate at the eavesdropper end:
\begin{equation}
\begin{aligned}
\frac{\partial C_{_E}}{\partial\textbf{f}_{s}^{*}} &=
\frac{P_{j}\textbf{H}_{je}^{*}\textbf{w}_{_E}\textbf{w}_{_E}^{*}\textbf{H}_{je}\textbf{f}_{j}}{\sigma_{_E}^{2} + P_{j} \Psi_{je} + P_{s} \Psi_{se}}.
\end{aligned}
\end{equation}
Thereby, we can express the gradient  of $\textbf{f}_{s}$ by:
\begin{equation}\label{gradientFs}
\begin{aligned}
\nabla_{\textbf{f}_{s}^{*}}C_{_s}=\frac{P_{j}\textbf{H}_{jl}^{*}\textbf{w}_{_L}\textbf{w}_{_L}^{*}\textbf{H}_{jl}\textbf{f}_{j}}{\sigma_{_L}^{2} + P_{j} \Psi_{jl} + P_{s} \Psi_{sl}}-\frac{P_{j}\textbf{H}_{je}^{*}\textbf{w}_{_E}\textbf{w}_{_E}^{*}\textbf{H}_{je}\textbf{f}_{j}}{\sigma_{_E}^{2} + P_{j} \Psi_{je} + P_{s} \Psi_{se}}.
\end{aligned}
\end{equation}

After computing the gradients, we can now move to the Gradient ascent algorithm. We presents two different algorithms. Algorithm 1 search how to maximize the secrecy capacity while maintaining the source power $P_{s}$ fixed. This is very efficient in case where the power is limited. Hence, the algorithm search for the perfect precoders/combiners configuration that will maximize the secrecy under power constraints. On the other hand, Algorithm 2 search for the best precoders/combiners combinations but with respect to a secrecy capacity target $\zeta$ required by the QoS. In this second approach, we fix $P_j$ and we search for the maximum possible secrecy capacity. If we achieve $\zeta$, the algorithm stops, otherwise it adapts $P_s$ while the power cannot exceed a certain threshold $\mu$ (to avoid infinite loop and unrealistic power  consumption). 
\begin{algorithm}
 \caption{\\Adaptive Gradient Ascent for fixed transmission power ($P_{j}$)}\label{algo}
 \begin{algorithmic}[1]
 \renewcommand{\algorithmicrequire}{\textbf{Input:}}
 \renewcommand{\algorithmicensure}{\textbf{Output:}}
 \renewcommand{\algorithmiccomment}{$\triangleright~$}
 \REQUIRE $\delta$, $\epsilon$, $\bold{H}_{sl},~\bold{H}_{se}$,~$\bold{H}_{jl}$,~$\bold{H}_{je}$,~ $\bold{w}_{_L}$ 
 \ENSURE $\bold{w}_{_L}$,~$\bold{f}_{s}$, ~$\bold{f}_{j}$\\
  \STATE \textbf{Initialize} $\bold{w}_{_L} = \bold{w}_{_L}^{(0)},~\bold{f}_s = \bold{f}^{(0)}_s, ~\bold{f}_j = \bold{f}^{(0)}_j$
  \WHILE{$|C_{s}^{(n+1)} - C_{s}^{(n)}| > \epsilon$ and $P_s \leq \mu$} 
    \STATE $C_{s}^{(n)} \gets C_{s}^{(n+1)}$
  \STATE $\bold{w}_{_L}^{(n+1)} \gets \bold{w}_{_L}^{(n)}+\delta \nabla_{\bold{w}_{_L}}C_{s}^{(n)}$
  \STATE $\bold{w}_{_L}^{(n+1)} \gets \frac{\bold{w}_{_L}^{(n+1)}}{\|\bold{w}_{_L}^{(n+1)}\|_2}$
  \COMMENT{Unit-norm constraint}
  \STATE $\bold{w}_{_L}^{(n+1)} \gets \frac{\bold{w}_{_L}^{(n+1)}}{\sqrt{N_R}|\bold{w}_{_L}^{(n+1)}|}$
  \COMMENT{CA constraint}
  \STATE $\bold{f}_j^{(n+1)} \gets \bold{f}_j^{(n)}+\delta \nabla_{\bold{f}_j}C_{s}^{(n)}$
  \STATE $\bold{f}_j^{(n+1)} \gets \frac{\bold{f}_j^{(n+1)}}{\|\bold{f}_j^{(n+1)}\|_2}$
  \STATE $\bold{f}_{j}^{(n+1)} \gets \frac{\bold{f}_{j}^{(n+1)}}{\sqrt{N_T}|\bold{f}_{j}^{(n+1)}|}$
  \STATE $\bold{f}_s^{(n+1)} \gets \bold{f}_j^{(n)}+\delta \nabla_{\bold{f}_s}C_{s}^{(n)}$
  \STATE $\bold{f}_s^{(n+1)} \gets \frac{\bold{f}_s^{(n+1)}}{\|\bold{f}_s^{(n+1)}\|_2}$
   \STATE $\bold{f}_{s}^{(n+1)} \gets \frac{\bold{f}_{s}^{(n+1)}}{\sqrt{N_T}|\bold{f}_{s}^{(n+1)}|}$
    \IF{$C_{s}^{(n+1)}<C_{s}^{(n)}$} \STATE {Adapt $\delta$} \ENDIF
  \ENDWHILE
 \RETURN $\bold{w}_{_L}$,~$\bold{f}_{s}$, ~$\bold{f}_{j}$, $C_{s}^{(n+1)}$
 \end{algorithmic} 
 \end{algorithm}
 
 \begin{algorithm}
 \caption{\\Adaptive Gradient Ascent for variable  transmission power ($P_{s}$)}\label{algo}
 \begin{algorithmic}[1]
 \renewcommand{\algorithmicrequire}{\textbf{Input:}}
 \renewcommand{\algorithmicensure}{\textbf{Output:}}
 \renewcommand{\algorithmiccomment}{$\triangleright~$}
 \REQUIRE $\zeta$, $\delta$, $\epsilon$, $\bold{H}_{sl},~\bold{H}_{se}$,~$\bold{H}_{jl}$,~$\bold{H}_{je}$,~ $\bold{w}_{_L}$ 
 \ENSURE $\bold{w}_{_L}$,~$\bold{f}_{s}$, ~$\bold{f}_{j}$\\
  \STATE \textbf{Initialize} $\bold{w}_{_L} = \bold{w}_{_L}^{(0)},~\bold{f}_s = \bold{f}^{(0)}_s, ~\bold{f}_j = \bold{f}^{(0)}_j$
  \WHILE{$C_{s}^{(n+1)} < \zeta$} 
  \STATE Call \textbf{Algorithm 1} from line \textbf{2} to line \textbf{16}
  \STATE  $P_s \gets P_s+\kappa P_s$
   \COMMENT{Adopt$ ~P_s$}
  \ENDWHILE 
 \RETURN $\bold{w}_{_L}$,~$\bold{f}_{s}$, ~$\bold{f}_{j}$, $C_{s}^{(n+1)}$
 \end{algorithmic} 
 \end{algorithm}
 \subsection{Gradient Learning Approach While Optimizing $W_{_E}$}
 To characterize the system performance, we used two different benchmarkers.
 The first one is computing the gradient of ${w}_{_E}$. It is clear that it is not feasible to design the combiner of the attacker. However, we can optimize with respect to ${w}_{_E}$, which of course will provide higher secrecy capacity, and we use the output result as a benchmarking tool. 
 The gradient of $\textbf{w}_{_E}$, which is given by
\begin{equation}
\begin{aligned}
&\nabla_{\textbf{w}_{_E}^{*}}C_{_s} =\frac{\partial C_{_s}}{\partial\textbf{w}_{_E}^{*}}= \left[ \frac{\partial C_{_L}}{\partial\textbf{w}_{_E}^{*}}-\frac{\partial C_{_E}}{\partial\textbf{w}_{_E}^{*}} \right]=-\frac{\partial C_{_E}}{\partial\textbf{w}_{_E}^{*}}\\
=&\frac{\textbf{w}_{_E}\sigma^2 + P_{j}\textbf{H}_{je}\textbf{f}_{j}\textbf{f}_{j}^{*}\textbf{H}_{je}^{*}\textbf{w}_{_E} }{\sigma_{_E}^{2} + P_{j} \Psi_{je} }\\
&-\frac{\textbf{w}_{_E}\sigma^2 + P_{j}\textbf{H}_{je}\textbf{f}_{j}\textbf{f}_{j}^{*}\textbf{H}_{je}^{*}\textbf{w}_{_E} +P_{s}\textbf{H}_{se}\textbf{f}_{s}\textbf{f}_{s}^{*}\textbf{H}_{se}^{*}\textbf{w}_{_E}}{\sigma_{_E}^{2} + P_{j} \Psi_{je}  + P_{s} \Psi_{se}}.
\end{aligned}
\end{equation}
After computing $\nabla_{\textbf{w}_{_E}^{*}}C_{_s}$, we optimize with respect to  $\textbf{w}_{_E}$ by implementing the following code line after the line with index $6$ in \textbf{Algorithm 1}:
$$
 \begin{cases} \bold{w}_{_E}^{(n+1)} \gets \bold{w}_{_E}^{(n)}+\delta \nabla_{\bold{w}_{_E}}C_{s}^{(n)} \\
\bold{w}_{_E}^{(n+1)} \gets \frac{\bold{w}_{_E}^{(n+1)}}{\|\bold{w}_{_E}^{(n+1)}\|_2}\\
\bold{w}_{_E}^{(n+1)} \gets \frac{\bold{w}_{_E}^{(n+1)}}{\sqrt{N_R}|\bold{w}_{_E}^{(n+1)}|}
\end{cases}
$$
 \subsection{SVD Upper Bound}
The second benchmarker is an upper bound by applying the singular value decomposition (SVD) on the channels $\textbf{H}_{sl}$, $\textbf{H}_{se}$, $\textbf{H}_{jl}$, and $\textbf{H}_{je}$.
Therefore, the upper bound of the achievable rate is expressed as follow
\begin{equation}\label{upper}
\begin{aligned}
&C_{s}^{\text{ub}} =\\ &\log_2\left(1 + \frac{P_{s}(\lambda_{11}^{sl})^{2}}{\sigma_{_L}^2 + (\lambda_{Nr Nt}^{jl})^{2}}\right) -  \log_2\left(1 + \frac{P_s(\lambda_{Nr Nt}^{se})^{2}}{\sigma_{_E}^2+P_j (\lambda_{11}^{je})^{2}}\right)  
\end{aligned}
\end{equation}
where $\lambda_{11}^{sl}$, and $\lambda_{11}^{je}$ are the maximum singular values of $\bold{H}_{sl}$, and $\bold{H}_{je}$, respectively.  $\lambda_{Nr Nt}^{se}$ and $\lambda_{Nr Nt}^{jl}$ are the minimum singular value of $\bold{H}_{se}$ and $\bold{H}_{jl}$, respectively. Although, the rate (\ref{upper}) is not achievable, it is a tool to verify the optimized system performances. 
 \section{Numerical Results and Discussions} 
\begin{table}[ht]
\caption{System Parameters}
\begin{tabular}{|c|c|c|}
\hline
\rowcolor[HTML]{FFCC67} 
Parameters                             & mmWave & Sub-6 GHz \\ \hline
Number of antennas at the jammer       & 64      & 16        \\ \hline
\rowcolor[HTML]{FFCC67} 
Number of antennas at the BS           & 64      & 16        \\ \hline
Number of antennas at the receivers    & 4       & 4         \\ \hline
\rowcolor[HTML]{FFCC67} 
Number of  clusters                    & 4       & 10        \\ \hline
Number of rays per cluster             & 15      & 20        \\ \hline
\rowcolor[HTML]{FFCC67} 
Angular spread                         & $10^{o}$       & $10^{o}$         \\ \hline
Power at the BS $P_s$                   & 10 dB   & 10 dB     \\ \hline
\rowcolor[HTML]{FFCC67} 
Power at the BS $P_s$                  & 10 dB   & 10 dB     \\ \hline
Step size $\delta$                                  & 0.1     & 0.1       \\ \hline
\rowcolor[HTML]{FFCC67} 
Convergence criterion $\epsilon$                                & $10^{-7}$    & $10^{-7}$      \\ \hline
Power adaptation rate $\kappa$                                  &   $10^{-2}$      &   $10^{-2}$    \\ \hline
\rowcolor[HTML]{FFCC67} 
Number of Monte Carlo simulation loops & 1000    & 1000      \\ \hline
\end{tabular}
\end{table}
In this section, we will examine the performance of the proposed security scheme.  To guarantee the convergence stability and to adjust any possible perturbation during the learning process, the step size parameter $\delta$ is initialized to 0.1 and it is divided by 2 whenever a perturbation occurs during the optimization cycle of the cost function. optimization.  The channel matrices $\bold{H}_{sl}$, $\bold {H}_{se}$, $\bold{H}_{je}$, the combiners $\bold{w}_{_L}$, $\bold{w}_{_E}$ and the precoders $\bold{f}_{s}$, and $\bold{f}_{j}$ are assumed to follow complex Gaussian distribution.
 
 \begin{figure}[ht]
\includegraphics[ width=\linewidth]{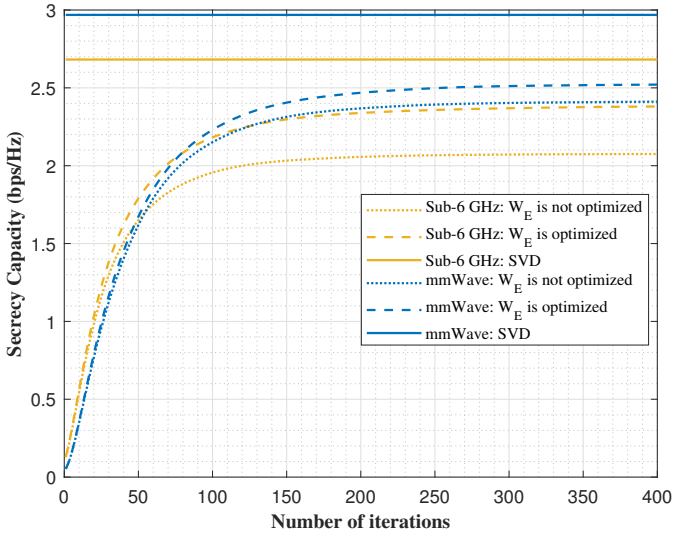}
\caption{Sub-6 GHz vs. mmWave: Comparison in terms of maximum achievable secrecy capacity and number of iterations under fixed source power $P_s$ }
\end{figure}
In Fig. 2, we studied the convergence rate of the optimization process for Sub-6 GHz and mmWave scenarios. We observe that the secrecy capacity for both scenarios converges with the same number of iterations which is 400 iterations. Regarding the achievable secrecy capacity (without optimizing $\bold{w}_{_E}$), Sub-6 GHz provides higher secrecy than the mmWave for the first 60 iterations. However, when the cost function starts to converge, mmWave offers 2.4 bps/Hz while Sub-6 GHz achieves about 2.1 bps/Hz. For the first benchmarker ( (while optimizing $\bold{w}_{_E}$), $\bold{f}_{s}$, and $\bold{f}_{j}$), we observe that this upper bound and the curve related to the case of random $\bold{w}_{_E}$ are very close for the first 100 iterations and a gap starts to grow when the learning process starts to converge.  We should note that the converged secrecy capacity did not exceed neither the first benchmarker, neither the second benchmarker( the SVD ceil) during the entire learning process.

\begin{figure}
\includegraphics[ width=\linewidth]{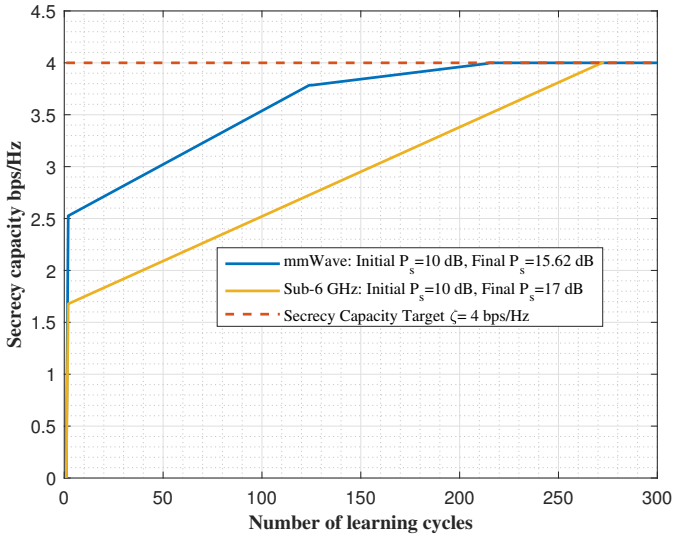}
\caption{Sub-6 GHz vs. mmWave: Comparison in terms of source power $P_s$  and of number of cycles required to reach a target secrecy capacity  $\zeta$. }
\end{figure}
For the previous Fig. 2, we discussed the results of Algorithm 1, where $P_s$ is fixed. Now, we move to analyze Algorithm 2 by discussing Fig. 3. In this scenario, the source power from the BS is considered to be variable. Our goal is to achieve a secrecy capacity noted by $\zeta$.  The algorithm searches for the maximum possible $C_{s}$ for a given $P_s$, if the obtained value reaches $\zeta$, the learning process will stop, if not, the source power will be adapted and another learning cycle will initiate (where a cycle is referring to the learning process for a given $P_s$). In Fig. 3, the target secrecy capacity is $\zeta$=4 bps/Hz while $P_s$ starts at 10 dB.  The algorithm arrives at the required $\zeta$ after 212 cycles for mmWave system, while it spends 272 cycles for Sub-6 GHz. Regarding the source power needed to optain $\zeta$, at the end of the entire learning process, $P_s$  increased from 10 dB to 15.62 dB for Sub-6 GHz while it requires only 17 dB in the case of mmWave. Therefore, under the condition of secrecy capacity requirement, we observe that the learning rate (number of cycles required to reach $\zeta$) for mmWave system  case is much higher than the Sub-6 GHz, while there is no significant difference concerning the needed source power $P_s$
\section{Conclusion}
In this work, we discussed the PLS for V2I communication under the scenario of eavesdropping attack for two different cases: Sub-6 GHz and mmWave systems. We assumed the deployment of a friendly jammer that aids to maximize the average secrecy capacity. We used adaptive Gradient ascent as a machine learning approach to maximize the secrecy capacity while searching for the best combination of combiner at the legitimate receiver and precoders at  BS/jammer.  We proposed two different algorithms.  In the first one,  we assumed that the source power is fixed and we searched for the best precoders and combiner that maximize the secrecy capacity. We noted that the cost function converges to the maximum secrecy capacity with the same rate for both the Sub-6 GHz and mmWave systems where the later achieved higher secrecy capacity than the former.
In the second algorithm, we suppose that the QoS requires a specific secrecy capacity threshold that we need to achieve while using the optimum source power. The mmWave system reached the secrecy target faster than the Sub-6 GHz.

\bibliographystyle{IEEEtran}
\bibliography{main}

\end{document}